\documentclass[reprint]{revtex4-1}
\usepackage{amsmath, amsfonts, amssymb, fullpage, graphicx}

\begin{document}
\title{Complex non--Hermitian Potentials and Real--Time Time--Dependent Density Functional Theory: A Master Equation Approach}
\author{Justin E. Elenewski$^1$, Yanxiang Zhao$^2$, and Hanning Chen$^1$}
\affiliation{$^1$Department of Chemistry and $^2$Department of Mathematics\\ The George Washington University, Washington, DC 20052.}
\email{chenhanning@gwu.edu}

\begin{abstract}
The simulation of quantum transport in a realistic, many--particle system is a nontrivial problem with no quantitatively satisfactory solution. While real--time propagation has the potential to overcome the shortcomings of conventional transport methods, this approach is prone to finite size effects that are associated with modeling an open system on a closed spatial domain. Using a master equation framework, we exploit an equivalence between the superoperators coupling an open system to external particle reservoirs and non--Hermitian terms defined at the periphery of a quantum device. By taking the mean-field limit, the equation of motion for the single--particle reduced density matrix becomes equivalent to real--time time--dependent density functional theory in the presence of imaginary source and sink potentials.  This method may be used to converge nonequilibrium steady states for a many--body quantum system using a previously reported constraint algorithm.
\end{abstract}

\maketitle

\section{Introduction}

\par While coarse--grained models are sufficient to describe the transport characteristics of many mesoscale and nanoscale systems, explicit electronic structure calculations are required to model realistic, sub--nanometer quantum devices.  The most effective theoretical framework utilizes density functional theory (DFT) \cite{Hohenberg1964,Kohn1965} in conjunction with the nonequilibrium Green's function (NEGF) formalism to determine the conductances and currents through these constrictions \cite{Stefanucci2004, DiVentra2000, Xue2001,  Xue2002}.  Unfortunately, these methods are only qualitative,  predicting currents that diverge from experiment by several orders of magnitude.  This discrepancy arises from the neglect of dynamic correlation effects within DFT as well as the inability of the DFT+NEGF method to describe charge reorganization during transport \cite{Gorling1995, Evers2004, Koentopp2008}.

\par An alternative approach, explicitly treating the electronic configuration of the system at every point in time, is to propagate the electronic wavefunctions using real--time time--dependent DFT (RT--TDDFT)  or Ehrenfest dynamics \cite{Yabana1996, Li2005}.  While adequately capturing dynamic phenomena, these calculations are performed within a closed simulation system.  As such, the method is  prone to unphysical particle accumulation at, or reflection from, the hard system boundaries \cite{Qian2006, Varga2011}.  These complications have limited the use of RT--TDDFT to extremely short simulation times and thus to a limited number of materials.

\par An emerging approach for the simulation of open quantum devices involves the use of imaginary, non--Hermitian potentials defined near the boundaries of a closed system \cite{Neuhasuer1989,Ferry1999, Muga2004}.  The full Hamiltonian is then of the form $\hat{H}' = \hat{H} + \hat{V}_\text{B}$, where $\hat{H}$ is the simulation Hamiltonian and  $\hat{V}_\text{B} = f(x) \pm i\Gamma(x)$ is a boundary term.  In this case both $f(x)$ and $\Gamma(x)$ are real functions that determine the spatial behavior of $\hat{V}_\text{B}$, with the provision that $\Gamma(x) \geq 0$ at all points.  An arbitrary state $\psi(x,t)$ will evolve under the influence of $\hat{H}$ so that $\psi(x,t') = \hat{U}(t',t) \psi(x,t) = \exp[-i \hat{H}(t'-t)] \psi(x,t)$, assuming that the system is translationally invariant in time (using natural atomic units $\hbar = m_e = e = 1$).  Since the time evolution operator $\hat{U}(t',t)$ is unitary,  the norm of the state $\vert \vert \psi(x,t') \vert \vert^2 = \vert \vert \psi(x,t) \vert \vert^2 = \int \psi^*(x,t) \psi(x,t) \, dV$ remains constant, reflecting conservation of the probability density $\rho(x,t) = \psi^*(x,t) \psi(x,t)$.  

\par In contrast, the non--Hermitian boundary term violates unitarity since $\rho(x,t') = \,\exp[\pm 2\Gamma(x) (t'-t)] \rho(x,t) $, leading to either an increase or a decrease in the norm of $\psi(x,t)$ depending on the sign of $\pm\Gamma(x)$.  In this manner, a potential with a positive imaginary component at all points may be treated as a `source' term for incoming particle density, while a potential with a negative imaginary term acts as a `sink' that emulates the loss of particles from the system.   This leads to a continuity equation 

\begin{multline} \label{continuity}
\frac{\partial}{\partial t} \int_{\mathcal{V}} \rho(x,t) \, dV +  \int_{\mathcal{V}} \nabla \cdot \vec{j}(x,t)\, dV\\ = 2 \int_{\mathcal{V}} \text{Im}[\hat{V}_B(x)] \rho(x,t)\,dV,
\end{multline}

\noindent where $\mathcal{V}$ is the volume on which $\text{Im}[\hat{V}_B(x)] \neq 0$ and $\vec{j}(x,t) =  \text{Im} [\psi^*(x,t) \nabla\psi(x,t)]$ is the probability current density.  The first term is associated with a growth in, or reduction of, the number of particles contained within $\mathcal{V}$, while the second term reflects the change in particle number within the volume $\mathcal{V}$ due to a flux through the boundary $\partial \mathcal{V}$.  The remaining term is dictated by the non--Hermitian behavior of  $\hat{V}_B(x)$.  It should be noted that this method only rescales the magnitude of a state vector $\psi(x,t) \mapsto \alpha \psi(x,t)$ for some $\alpha \in \mathbb{R}^+$.  As a consequence, it cannot alter the Fock space occupancy of a theory (except by incidentally setting the magnitude of some $\psi(x,t)$ to zero during numerical evaluation).  This can, however, lead to additional unphysical behavior in a fermionic system if $\alpha > 1$, and hence care must be taken to ensure that $0 < \alpha \leq 1$ in any algorithmic method \cite{Elenewski2014}.

\par This approach is highly efficacious when treating the loss of particles from a system, with applications ranging from simulations of wavepacket scattering and resonance states to transport and nuclear tunneling \cite{Moiseyev1998,Muga2004, Varga2007, Varga2009, Varga2011, Moiseyev2011, Scamps2015}. As a consequence, the properties of sink potentials have been exhaustively optimized to maximize particle absorption and minimize spurious reflections \cite{Neuhasuer1989, Vibok1992, Vibok1992b, Brouard1994, Riss1996, Ge1997, Palao1998, Ferry1999, Manolopoulos2002, Poirier2003, Poirier2003b, Muga2004}.  In fact, this method has recently been conjoined with RT--TDDFT to afford simulations of electronic transport with no constraints other than those inherent to DFT \cite{Varga2011}.  Unfortunately, the persistent removal of particles through the absorbing boundary will ultimately retard time--dependent transport, leading to an attenuation of current characteristics due to the progressive ionization of the system.  Furthermore, attempts to balance particle loss with a compensatory source potential generally fail or require elaborate algorithmic constructions \cite{Wibking2012} and manual tuning.  
\par In an effort to circumvent this limitation, the  authors  recently introduced a numerical protocol to converge the nonequilibrium steady--states (NESS) of open quantum systems through the use of these imaginary boundary potentials \cite{Elenewski2014, Elenewski2015}.   When conjoined with an electronic structure method \cite{Hohenberg1964,Kohn1965}, this technique affords a promising framework for the simulation of transport in technologically relevant materials at quantitative accuracy.  This construction is nonetheless heuristic and its is not rigorously defined within the context of the  many--body problem.  To provide this justification, it is demonstrated herein that our algorithm is associated with the mean--field limit of a general electronic Hamiltonian.  Using a designated series of approximations, this limit may be taken to coincide with RT--TDDFT as well as higher--order treatments of  correlation based on the DFT + $U$ method \cite{Anisimov1991, Anisimov1991b}.

Our approach is as follows.  We will review a framework for an arbitrary open field theory with two--body interactions and demonstrate its association with a suitable Lindblad--like master equation.  We will then analyze the evolution of the reduced two--particle density matrix in the context of a discrete, lattice--based model that encapsulates several electronic structure methods.  From here, we will demonstrate that the mean--field limit corresponding to RT--TDDFT is accommodated through our existing simulation protocol for open systems, provided that the non--Hermitian gain and loss terms are balanced \cite{Elenewski2015}.  While higher--order solutions of our master equation may be obtained using a variational approach \cite{Weimer2015}, we are content with the mean field case due to its correspondence with widely used simulation methods. Finally, we perform a series of RT--TDDFT tests to demonstrate the numerical performance and stability of our method for a toy multiparticle system.
\section{Continuum Field Theory for Transport}

\begin{figure}
\includegraphics[scale=0.85]{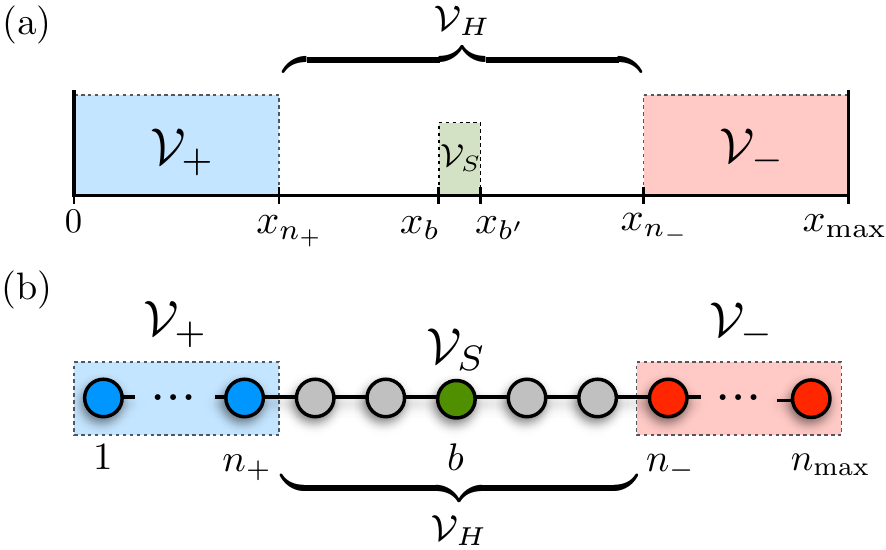}
\caption{(a) Continuum model of a quantum device modeled as a Hermitian component $\mathcal{V}_H$ situated between imaginary source and sink potentials in the incoming $\mathcal{V}_+$ and outgoing $\mathcal{V}_-$ leads.  The Hermitian domain contains a scattering center $\mathcal{V}_S$ responsible for the device characteristics. (b) Discrete model of the same system comprising a finite set of lattice sites.}
\end{figure}

\par To establish a formal correspondence between an open system described using quantum master equations  and a description in terms of complex boundary potentials, it is useful to consider a generic theory of interacting  fermions.  This model reflects the `bulk' of a quantum device, with electronic transport occurring through the exchange of particles with infinite `leads' (reservoirs) at situated at the device boundaries (Fig. 1a).  In a real space picture, the essential physics is encapsulated in a Hamiltonian $H' = H + V_B$, with $V_B$ a non--Hermitian boundary potential term and $H$ defined so that 

\begin{multline} \label{ftHamiltonian}
H = \int_\mathcal{B} dx \, \psi^\dagger (x) h_0 (x) \psi(x) + \\
\frac{1}{2} \int_\mathcal{B} dx \, dx' \, \psi^\dagger (x) \psi^\dagger (x') U(x,x') \psi(x') \psi(x).
\end{multline}

\noindent Here  $h_0(x) = -\nabla^2/2$ is a `kinetic' operator and $U(x,x')$  is a two--body potential.  By convention, the fermionic field operators $\psi^\dagger (x), \psi(x)$ satisfy canonical anticommutation relations $\{\psi(x), \psi(x')\} =0$ and $\{\psi(x), \psi^\dagger (x')\} = \delta(x-x')$ where $\delta(x - x')$ is the Dirac delta function.  Following this construction, the domain $\mathcal{B}$ is a compact submanifold $\mathcal{B} \subset \mathbb{R}^3$ on which $\psi(x)$ and $\psi^\dagger(x)$ are supported.  Furthermore, there exists a Hermitian subdomain $\mathcal{V}_H \subset \mathcal{B}$ corresponding to the interior of the device, on which $V_B$ vanishes.

\par The device terminals are modeled by using imaginary source $iV_+$ and sink $iV_-$ potentials, defined on nonintersecting domains $\mathcal{V}_\pm \subset \mathcal{B}$  that are situated at the incoming and outgoing  boundaries.  This corresponds to a two--terminal system, however, the arguments herein may be expanded to arbitrary number of contacts without modification.  Under these provisions, the boundary potentials assume a simple representation in terms of the fermionic fields

\begin{equation} \label{cplxPotForm}
iV_{\pm} = \pm i\int_{\mathcal{B}} dx \, \psi^\dagger(x) \Gamma^\pm(x) \psi(x),
\end{equation}

\noindent where $\Gamma^\pm (x)$ is the potential strength. From this construction, it follows that $V_B = i(V_+ + V_-)$, completing the model. Assuming that $\Gamma^\pm(x) \geq 0$, the positive sign corresponds to a particle source, and the negative sign to a particle sink. Within a certain limit, this theory admits an identical description in terms of Lindbladian dynamics \cite{Selsto2010, Selsto2012}.

\par The Lindblad formalism \cite{Lindblad1976, Gorini1976, Schaller2014} describes the evolution of an open quantum system, with density matrix $\hat{\rho}$, through a Markovian master equation

\begin{equation} \label{themastereqn}
i\frac{d}{dt} \hat{\rho} = [\hat{H}, \hat{\rho}] + \mathcal{L}(\hat{\rho}),
\end{equation}

\noindent where the action of the Lindbladian superoperator $\mathcal{L}$ is defined as

\begin{equation} \label{GenLindbladian}
\mathcal{L}(\hat{\rho}) = i \sum_{jk} \gamma_{jk} (\hat{A}^\dagger_j \hat{A}_k \hat{\rho} + \hat{\rho}\hat{A}^\dagger_j \hat{A}_k - 2 \hat{A}_j \hat{\rho} \hat{A}^\dagger_k).
\end{equation}

\noindent In this case $\hat{A}, \hat{A}^\dagger$ are termed Lindblad operators and are required to constitute a Krauss map $\sum_{jk} \gamma_{jk} \hat{A}_j \rho \hat{A}_k^\dagger \longrightarrow \rho'$ that preserves the trace and positivity of the density operator, while the coefficients $\gamma_{jk}$ constitute a positive--semidefinite matrix \cite{Schaller2014}.  Note that in the absence of the $\mathcal{L}(\hat{\rho})$ term, Eq. (\ref{themastereqn}) is simply the Liouville--von Neumann equation.

\par To maintain consistency between the master equation and non--Hermitian descriptions, the imaginary terms contained within $V_B$ must be mapped to an equivalent set of Lindblad operators.  This connection is straightforward after noting that the isolated boundary term $V_-$ is associated with a Liouville--von Neumann equation

\begin{equation}
\begin{split}
i\frac{d}{dt} \hat{\rho} &= \hat{H}' \hat{\rho} - \hat{\rho} (\hat{H}')^\dagger \\
&=  [\hat{H}, \hat{\rho}] + i(\hat{V}_- \hat{\rho} + \hat{\rho} \hat{V}_- ).
\end{split}  
\end{equation}

\noindent This has a natural correspondence with Eq. (\ref{GenLindbladian}) provided the Lindblad operators satisfy

\begin{equation}
V_- = \int_\mathcal{B} dx \, \hat{A}^\dagger (x) \hat{A}(x),
\end{equation}

\noindent with the added convenience of being a diagonal representation.  To solidify this correspondence, it is also necessary to require that the term $-2 \hat{A}_n \hat{\rho} \hat{A}^\dagger_n$ maps an $N$--particle system to an $(N-1)$--particle system and that the adjoint term $-2 \hat{A}^\dagger_n \hat{\rho} \hat{A}_n$ maps an $N$--particle system to an $(N+1)$--particle system \cite{Selsto2010,Selsto2012}.   These latter contributions are only relevant when the particle number of a given state changes by a full unit; that is, when a state is initially created or ultimately annihilated.

 The simplest construction satisfying this requirement is of the form $\hat{A}(x) = \sqrt{\Gamma(x)}\, \psi(x)$, leading to a pair of Lindbladians
 
\begin{multline}
\mathcal{L}_- (\rho) = -i\int_\mathcal{B} dx\, \{\Gamma^-(x) \psi^\dagger(x) \psi(x), \rho(x)\} \\ + 2i \int_\mathcal{B} dx \, \Gamma^-(x) \psi(x) \rho(x) \psi^\dagger (x) 
\end{multline}
\begin{multline}
\mathcal{L}_+ (\rho) =  i\int_\mathcal{B} dx\, \{\Gamma^+(x) \psi(x) \psi^\dagger(x), \rho(x)\} \\ - 2i \int_\mathcal{B} dx \, \Gamma^+(x) \psi^\dagger(x) \rho(x) \psi (x)
\end{multline}

\noindent and an expanded master equation

\begin{equation} \label{masterEqnBoth}
i \frac{d \rho}{d t} = [H,\rho] + \mathcal{L}_- (\rho) + \mathcal{L}_+ (\rho).
\end{equation}

\noindent This is just the conventional Lindblad representation for the coupling of an open system to a pair of particle reservoirs.  As such, the use of complex source and sink potentials may be taken as equivalent to Lindbladian dynamics if the norm changes by less than a full unit for a given fermionic state, affording a formal justification for the use of this method to emulate open quantum systems \cite{Selsto2010, Selsto2012}.  These restrictions are conveniently satisfied by nonequilibrium steady--states.

\par In subsequent developments, the imaginary potentials will be defined as piecewise functions 

\begin{equation}
i\Gamma^\pm (x) = \begin{cases}
      i\Gamma_0^\pm & x \in \mathcal{V}_\pm, \\
      0 & \text{otherwise} 
   \end{cases}
\end{equation}

\noindent where $\Gamma_0^\pm \in \mathbb{R}$ is a constant.  In this case, the potential term (Eq. (\ref{cplxPotForm})) assumes a particularly simple form

\begin{equation}
iV_\pm = \pm (i\Gamma_0^\pm) \int_{\mathcal{V}_\pm} dx \, \psi^\dagger(x) \psi(x)
\end{equation}

\noindent and thus is only a function of the total density contained within the volume $\mathcal{V}_\pm$. This fact is of substantial algorithmic importance \cite{Elenewski2015}.

\section{Discrete Representation and Model Hamiltonains}

\par The continuum field theory developed above captures the essential physics associated with an open transport problem, however, a discrete representation must be assumed for practical calculations.  To this end, consider a one--dimensional $N$--particle system discretized on a lattice of spacing $\Delta x = x_{j+1} - x_{j}$ with a total of $n_\text{max}$ sites (Fig. 1b).  A convenient many--body Hamiltonian is of the form

\begin{multline}\label{genham}
\hat{H} = -\sum_{\langle jk \rangle,\sigma} (\eta_{jk}\, \hat{c}^\dagger_{j\sigma} \hat{c}_{k\sigma} + \eta_{kj}^* \, \hat{c}^\dagger_{k\sigma} \hat{c}_{j\sigma})\, + \\
\frac{1}{2} \sum_{\substack{\langle jk \rangle \\ \sigma,\sigma'}} \hat{c}^\dagger_{j\sigma} \hat{c}^\dagger_{k\sigma'} \, U_{jk} \, \hat{c}_{k\sigma'} \hat{c}_{j\sigma} +  \sum_{j,\sigma} \mu_j \, \hat{c}^\dagger_{j\sigma}\hat{c}_{j\sigma},
\end{multline}

\noindent where the two--body potential $U_{jk} = U(x_j - x_k)$ is defined as a translationally invariant function, the hopping terms $\eta_{jk}$ corresponding to the intersite coupling amplitudes, and $\sigma$ is a spin index.  The discrete creation and annihilation operators now satisfy the anticommutation relations $\{\hat{c}_{j\sigma}, \hat{c}_{k\sigma'}\} = 0$ and $\{\hat{c}_{j\sigma}, \hat{c}_{k\sigma'}^\dagger\} = \delta_{jk} \delta_{\sigma \sigma'}$ where $\delta_{jk}$ is the Kronecker delta function.  These approximations are representative of most transport problems, and many two-- and three--dimensional systems become effectively one--dimensional when transport occurs along a single axis of propagation.

\par The model in Eq (\ref{genham}) is quite general.  If the intersite hopping in the first term is restricted to nearest--neighbor pairs $\langle j,k\rangle$ then this contribution is identical to the tight--binding model. Furthermore, if the two--body interaction is subsequently restricted to an on--site potential $U_{jk} = U_j \, (1 - \delta_{\sigma \sigma'}) \, \delta_{jk} $ associated with  spin--state occupancy, then the Hamiltonian $\hat{H}$ is that of the Hubbard model.  The remaining term is associated with an on--site chemical potential $\mu_j$ that may either generate a current or control the filling factor of the Hubbard Hamiltonian \cite{Hubbard1963, Fradkin1998}.   When applied to simple systems, including the noninteracting tight--binding model, the Lindblad framework has been shown to capture the physics of a dissipative nanoscale constriction or  junction \cite{Selsto2010, Dzhioev2011, Dzhioev2012, Medvedyeva2015} in a manner comparable to the Landauer formalism \cite{Landauer1957, Landauer1970, Fisher1981, Buttiker1985, Buttiker1986}. Furthermore, this particular approach has proven efficacious for modeling mean--field transport behavior in a double--well Bose--Einstein condensate coupled to external reservoirs, as derived from a Bose--Hubbard Hamiltonian \cite{Tikhonenkov2007, Trimborn2008, Witthaut2011, Single2014, Dast2014}

\par A few abbreviations and approximations are necessary to simplify discussion. Specifically, it is convenient to assume an identical coupling $\eta = \eta_{j,j+1} = \eta^*_{j+1,j}$ for left--moving and right--moving particle exchange between lattice sites.  Furthermore, to incorporate an externally applied electric field of magnitude $E$ it is convent to set $\mu_j = j E\,\Delta x$, where $\Delta x$ is the lattice spacing in the direction of the field. This affords the gradient required to generate a particle current.  When $U_{jk} = 0$, this particular model has been demonstrated to describe transport in open nanoscale wires under the influence of both weak and strong dissipation \cite{Medvedyeva2013, Medvedyeva2014, Medvedyeva2015}.  Finally, for pedagogical transparency, spin degrees of freedom will be ignored.

\par To characterize transport within the mean--field limit, it is necessary to monitor the time--evolution of the reduced one--particle density matrix $\tilde{\rho}^{(1)}_{jk}(t) = \langle \hat{c}_j^\dagger \hat{c}_k \rangle = \text{Tr} (\hat{c}_j^\dagger \hat{c}_k \hat{\rho}(t))$, where the trace is taken over all eigenstates of the Hamiltonian.  This approach has been rigorously explored for both open Bose--Hubbard dimers and chains \cite{Witthaut2011,Dast2014} and knowledge of this quantity is sufficient to calculate the currents within a nanoscale system \cite{Tu2008,Jin2010}.  As a first step, introduce the discrete superoperators

\begin{equation}
\mathcal{L}_-(\hat{\rho}) = -i \sum_j \{\Gamma^-(x_j) \hat{c}_j^\dagger \hat{c}_j, \hat{\rho}\} + 2i \sum_j \Gamma^-(x_j) (\hat{c}_j \hat{\rho} \hat{c}_j^\dagger ),
\end{equation}

\begin{equation}
\mathcal{L}_+(\hat{\rho}) = i \sum_j \{\Gamma^+(x_j)\hat{c}_j \hat{c}_j^\dagger, \hat{\rho}\} - 2i \sum_j \Gamma^+(x_j) (\hat{c}^\dagger_j \hat{\rho} \hat{c}_j ),
\end{equation}

\noindent with the provision that $\Gamma^\pm(x_j) = \Gamma_0^\pm$ for $x_j \in \mathcal{V}_\pm$ and $\Gamma^\pm(x_j) = 0$ otherwise. The master equation (\ref{masterEqnBoth}) uniquely  determines an equation of motion for $\tilde{\rho}^{(1)}_{jk}(t)$ as

\begin{equation}
\begin{split}
i \frac{\partial}{\partial t}\tilde{\rho}^{(1)}_{jk}(t) =& i\text{Tr}(\hat{c}^\dagger_j \hat{c}_k \dot{\hat{\rho}}(t))\\
=& \text{Tr} \left(\hat{c}_j^\dagger \hat{c}_k [\hat{H}, \hat{\rho}]\, + \right.  \left.\hat{c}^\dagger_j \hat{c}_k (\mathcal{L}_- (\hat{\rho}) + \mathcal{L}_+ (\hat{\rho}))\right) \\
=& \text{Tr} \left([\hat{c}_j^\dagger \hat{c}_k, \hat{H}]\hat{\rho}\, \right) + \\
&  \text{Tr}\left(\hat{c}^\dagger_j \hat{c}_k (\mathcal{L}_- (\hat{\rho}) + \mathcal{L}_+ (\hat{\rho}))\right),
\end{split}
\end{equation}

\noindent where the cyclic permutativity of the trace has been exploited to rearrange the commutator.   Using the model Hamiltonian (\ref{genham}), it is possible to expand this explicitly 

\begin{equation}
\begin{split}
i \frac{\partial}{\partial t}\tilde{\rho}^{(1)}_{jk}(t) =& \, \eta \left(\tilde{\rho}^{(1)}_{j+1,k} + \tilde{\rho}^{(1)}_{j-1,k} - \tilde{\rho}^{(1)}_{j,k+1} - \right. \\ 
& \left. \tilde{\rho}^{(1)}_{j,k-1}\right) - \frac{(\mu_j - \mu_k)}{2} \tilde{\rho}^{(1)}_{jk} \\
& + i(\Gamma^+_j + \Gamma^+_k) \tilde{\rho}^{(1)}_{jk}(t)  \\
& - i(\Gamma^-_j + \Gamma^-_k) \tilde{\rho}^{(1)}_{jk}(t) + \mathcal{K}[\hat{\rho}(t)],
\end{split}
\end{equation}

\noindent so that the term $\bar{\mathcal{K}}[\hat{\rho}(t)]$ is defined as

\begin{multline}
\mathcal{K}[\hat{\rho}(t)] =\frac{1}{2} \text{Tr} \left[ \hat{c}_j^\dagger \hat{c}_k  \left(\sum_{\substack{\langle mn \rangle}} \hat{c}^\dagger_{m} \hat{c}^\dagger_{n} \, U_{mn} \, \hat{c}_{n} \hat{c}_{m}\right) \hat{\rho} \right] - \\
\frac{1}{2} \text{Tr} \left[ \left(\sum_{\substack{\langle mn \rangle}} \hat{c}^\dagger_{m} \hat{c}^\dagger_{n} \, U_{mn} \, \hat{c}_{n} \hat{c}_{m}\right) \hat{c}_j^\dagger \hat{c}_k \hat{\rho} \right],
\end{multline}

\noindent and describes the many--particle interactions inherent to the system. This term is not tractable without the use of further approximations, as the equation of motion for each constituent density matrix contains a density matrix of higher order, forming the infinite Bogoliubov--Born--Green--Kirkwood--Yvon (BBGKY) hierarchy \cite{Fesciyan1973}.

\par To mimic the construction of DFT, it is necessary to introduce an effective potential  $\mathcal{K}[\hat{\rho}(t)] \mapsto \bar{\mathcal{K}}[\tilde{\rho}(t)] (\tilde{\rho}_{jk}^{(1)}(t) - \tilde{\rho}_{kj}^{(1)}(t))$ using the Hartree approximation. In doing so, the interaction potential of a many--body system is mapped onto a mean--field theory of non--interacting particles.   Unfortunately, a potential constructed in this manner is not guaranteed to satisfy the semigroup property required by the Lindblad theorem due to its explicit time dependence.  Despite this shortcoming, an identical master equation may be derived under the conditions at hand \cite{Burke2005}.   Since the time dependence of the potential is carried only by $\rho(t)$,  a given single--particle potential is in a one--to--one correspondence with the density if the functional form of the superoperators $\mathcal{L}_\pm (\rho)$ and electron--electron interaction are fixed.   This set of conditions is inherent to our construction, and affords a framework that parallels the construction of RT--TDDFT.

\par The final approximation is to introduce a mean--field ansatz \cite{Witthaut2011} for a single particle state $\tilde{\rho}^{(1)}_{jk} = \psi_j^* \psi_k$ where $\psi_j, \psi_k \in \mathbb{C}$ are complex amplitudes.  This affords an evolution equation for $\psi_k$ given by

\begin{multline}\label{singleparticle}
i \frac{d}{dt} \psi_k = -\eta (\psi_{k+1} + \psi_{k-1}) + \bar{\mathcal{K}}(t) \psi_k \\ 
+ \frac{\mu_k}{2} \psi_k + i (\Gamma^+_k - \Gamma^-_k)\psi_k,
\end{multline}

\noindent which is just the Schr\"{o}dinger equation for a single particle with wavefunction $\psi_k$, moving in the effective potential $\bar{\mathcal{K}}(t)$, and accompanied by boundary sources and sinks.  This equation of motion is identical to that of RT--TDDFT, implying that RT--TDDFT calculations can simulate the non--equilibrium steady--states of open quantum systems if they are augmented with appropriate non--Hermitian boundary terms.  

\par To make practical use of this result, define an effective Hamiltonian $\hat{H}_\text{eff}(t) = -\eta \,\partial_x^2 + \bar{\mathcal{K}}(t)  + \frac{\mu_k}{2}$ for the first three terms on the right--hand side of Eq. (\ref{singleparticle}).  It is assumed that this time--dependent Hamiltonian is piecewise constant $\hat{H}_\text{eff}(t) \approx \hat{H}_\text{eff}$ over some small interval $\delta t = (t' - t)$, as in most RT--TDDFT simulations.   The evolution equation  then has a formal solution in terms of $\hat{H}_\text{eff} $ so that
\begin{multline}
\psi_k(t') = \exp[-i\hat{H}_\text{eff} (t'-t)] \\ \exp [(\Gamma^+_k - \Gamma^-_k)(t'-t)] \, \psi_k(t),
\end{multline}

\noindent where the second exponential factor is associated with the non--unitary modulation of $\psi_k(t)$ during temporal propagation.  Considering only the non--unitary part, the density evolves as $\rho_k (t') = \psi^*_k(t')\psi_k(t') = \exp [2(\Gamma^+_k - \Gamma^-_k)(t' - t)] \rho_k(t)$ and hence the occupancy of each site is defined by the balance between $\Gamma^+_k $ and $\Gamma^-_k$.   The norm is then given by 

\begin{equation}
\begin{split}
\frac{\partial}{\partial t'}\mathcal{N}(t') & = \frac{\partial}{\partial t'} \sum_{k=1}^{n_\text{max}} \rho_k(t')\\
&=\frac{\partial}{\partial t'} \sum_{k=1}^{n_\text{max}} \exp [2(\Gamma^+_k - \Gamma^-_k)(t' - t)] \rho_k(t) \\
&= 2\sum_{k=1}^{n_+} \Gamma^+_k \exp[2 \Gamma^+_k  (t' - t)] \rho_k(t)  \\
&-2  \sum_{k=n_-}^{n_\text{max}} \Gamma^-_k \exp[-2 \Gamma^-_k (t' - t)] \rho_k(t), \\
\end{split}
\end{equation}

\noindent where $n_+$ and $n_-$ denote the terminal and initial points of the source and sink domains and $n_\text{max}$ is the number of lattice points (Fig. 1b).  If $\Gamma_k^+ = \Gamma_k^- = \Gamma_0$ is a fixed constant on each site, this expression simplifies considerably

\begin{multline}
\frac{\partial}{\partial t'}\mathcal{N}(t') = 2\Gamma_0 \left[ \sum_{k=1}^{n_+}\rho_k(t') - \sum_{k=n_-}^{n_\text{max}}\rho_k(t')\right], 
\end{multline}

\noindent identical to the continuum limit \cite{Elenewski2015}. Under these conditions, the evolution of $\rho(t)$ is contingent on the  net density contained within the source and sink regions.  If we introduce a constraint so that these sums are matched at every timestep, then $\partial_{t'} \mathcal{N}(t') = 0$ and the particle number in the system is constant. 

\par This construction is identical to a method previously reported by the authors, in which imaginary boundary conditions are employed to converge nonequilibrium steady--states for single particle transport in an open quantum system \cite{Elenewski2015}.  The extension reported above demonstrates that this method extrapolates to the mean--field, many--particle case, provided that the physics is appropriately described using a DFT potential $\bar{\mathcal{K}}(t) = \bar{\mathcal{K}}[\rho(t)]$.  Furthermore, in the context of an RT--TDDFT calculation, a constraint protocol must be used to individually balance the density within source and sink regions for each constituent state $\psi_\alpha$ at every timestep of propagation.  That is, the boundary conditions must be applied independently to each particle propagating between source and sink domains.

\section{Numerical Methods}

\par To demonstrate the efficacy of our constraint--based transport framework, RT--TDDFT calculations were performed using a two--particle system in which Hartree and exchange contributions are included in the effective potential.  A bias was applied across the system by fixing inhomogeneous boundary conditions for the Poisson solver, and a static potential barrier was introduced to act as a scattering center.   While numerical parameters for these simulations are outlined in this section, the algorithmic implementation is documented in the appendix to maintain brevity.  The choice of parameters is intended to rapidly demonstrate the capabilities of our method, however, an adaptation to more physically realistic conditions is straightforward.  

\par Simulations were performed on a one--dimensional domain, consisting of the interval $[0.0 \, a_0, 8.0\, a_0]$, with hard--wall boundary conditions applied at the limits of the system.  Here $a_0$ denotes the Bohr radius.  Imaginary source $iV_+ = i\Gamma^+_0$ and sink $iV_- = -i\Gamma^-_0$ potentials were defined on $\mathcal{V}_+ = (0.0\, a_0, 0.05\, a_0]$ and $\mathcal{V}_- = [7.9\, a_0, 8.0\, a_0)$, with potential strengths of identical magnitude in either region $\Gamma^+_0 = \Gamma^-_0 = 3.5 \times 10^4$ Ha (0.68 MeV) that differ only by sign.  A rectangular scattering barrier of magnitude $V_0 = 1.0 \times 10^5$ Ha (2.76 MeV) was situated on $\mathcal{V}_S = [a,b] = [3.9\, a_0 , 4.1\, a_0]$ and an energetic mismatch of $2.0$ Ha (54.4 eV) was applied to the electrostatic potential across the simulation. Real--time propagation was performed using a spatial discretization of $\Delta x = 1 \times 10^{-3}\, a_0$  and a timestep of $\Delta t = 5 \times 10^{-8} $ a.u. $ = 1.0 \times 10^{-28}$ s alongside a Runge--Kutta algorithm to integrate the equations of motion.

\par The initial wavefunctions were chosen to be time--independent scattering states for the rectangular barrier 

\begin{equation}
\psi_\alpha(x, t = 0) =  \begin{cases}
e^{i q_0 x} + B e^{-i q_0 x} & x \leq a \\
C e^{iq_0' x} + D e^{-i q_0' x} & a \leq x \leq b \\
 F e^{iq_0 x} & x \geq b,
\end{cases}
\end{equation}

\noindent where the initial wavevector $q_0 = 1000 \, a_0^{-1}$ is associated with an energy $E = q_0^2 / 2$ and where the wavevector in the scattering region $\mathcal{V}_S$ is $ q_0' = \sqrt{2(E - V_0)}$.  The density in the source region $\mathcal{V}_+$ was constrained every $N_R = 250$ timesteps by setting the wavefunction in $\mathcal{V}_+$ to $\psi(x,t_k)\vert_{x \in \mathcal{V}_+} = [\mathcal{N}_- (t_{k-1}) + \mathcal{N}_R (t_{k-1})]^{1/2} \psi_G(x,0)$, where $\psi_G(x,0) =  (\ell_+)^{-1/2} e^{i q_0 x}$ is a seed wavefunction and $\ell_+$ is the width of the generating region.  Here $\mathcal{N}_- (t_{k-1})$ is an estimator for the total density absorbed in $\mathcal{V}_-$ at the previous timestep and  $\mathcal{N}_R (t_{k-1})$ is an estimator for the total density reflected back into $\mathcal{V}_+$.  This situation simplifies considerably for the one--dimensional, two--terminal case (Fig. 1(a)), in which $\mathcal{N}_- (t_{k-1}) + \mathcal{N}_R (t_{k-1}) = 1 -  \mathcal{N}_H (t_{k-1})$, where $\mathcal{N}_H (t_{k-1})$ is the total density contained in the Hermitian region $\mathcal{V}_H = \mathcal{B} \setminus (\mathcal{V}_+ \cup \mathcal{V}_-)$ situated between the source and sink.  Note that this estimator is a first--order approach, and more inventive schemes are likely to lend improved convergence and greater numerical stability.  

\par Once the simulation has converged to a nonequilibrium steady state, charge accumulation or depletion will not occur at any point within the device.  Assuming a time--independent bias, this implies that the probability density fluxes observed before ($J_<$) and after ($J_>$) the scattering center will become equal.  Consequently, the ratio $ J_> / J_<$ will approach unity, affording an effective metric for convergence.  These fluxes were accordingly measured within volumes located before ($[3.45\, a_0, 3.60 \, a_0]$) and after ($[4.40\, a_0, 4.55\, a_0]$) the scattering barrier.  

\begin{figure}\label{scatterfigure}
\includegraphics[scale=1.0]{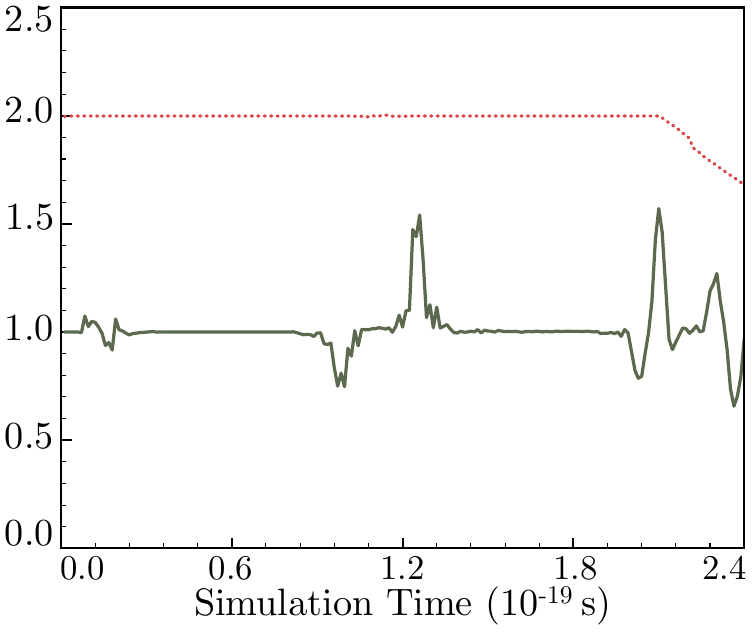}
\caption{Convergence of transport properties for two--particle scattering from a rectangular barrier, including the total norm (red, dashed line) and flux ratio $J_> / J_<$ (green, solid line).}
\end{figure}

\par The density constraint algorithm was applied for a total of $1.75 \times
10^5$ timesteps ($2.1 \times 10^{-19}$ s) during benchmark simulations, after which the constraint was released and strength of the source potential set to zero. The sink potential was left active to remove density from the system.  Under these conditions, the two--particle simulations exhibit excellent norm conservation and stable nonequilibrium steady states, with the norm remaining within 99.9\% of its target value (Fig. 2) throughout the calculation.  The current distribution is also highly stable, fluctuating by no more than 0.001\% between $3.6 \times 10^{-20}$ s and $8.2 \times 10^{-20}$ s, and by no more than 0.008\% between $1.5 \times 10^{-19}$ s and $1.9 \times 10^{-19}$ s.   These intervals are long enough for a unit of density moving at the group velocity $v_g$ to traverse half of the simulation, and thus exceed the relative timescale required for an accurate determination of currents in a transport calculation.

\par  The spikes and dips observed in $J_> / J_<$ are an artifact of the initial simulation conditions.  The early fluctuations (between $7.3 \times 10^{-21}$ s and $2.5 \times 10^{-20}$ s)  are associated with multiple reflections from the rectangular barrier that were not accommodated when choosing the starting wavefunction.  The second cluster (between $9.1 \times 10^{-20}$ s and $1.4 \times 10^{-19}$ s)  arises from a distortion near the source and sink regions that occurs early in the simulation. This modulation is a direct consequence the differing amplitudes of left-- and right--moving plane wave components.  This same issue manifests again at $2.0 \times 10^{-19}$ s, and repeats in a periodic yet damped manner if the simulation timescale is extended.  The use of a higher--order constraint / estimator algorithm for the generating region, as well as a more suitable choice of initial wavefunction, should remedy these issues in a realistic calculation. 

\par Finally, the trapping of density in bound states was assessed by observing the norm of the system at $5.0 \times 10^5$ timesteps ($6.0 \times 10^{-19}$ s),  with source potential and  constraint algorithm having been disabled at $1.75 \times 10^5$ timesteps ($2.1 \times 10^{-19}$ s).  In contrast to the single--particle case, less than 0.5\% of the probability density was retained within the system.  In prior calculations this accumulation was nontrivial and was associated with the scattering of spectral components to low wavevector modes \cite{Elenewski2015}.   It is expected that this quantity will be even smaller when a greater number of particles are present.   
\section{Conclusions}

\par The derivation presented herein affords a formal justification for the use of imaginary source and sink potentials when emulating open systems during real--time, many--body simulations.  If the appropriate approximations are made, this framework may be applied to RT--TDDFT calculations, where nonequilibrium steady--states are readily converged by balancing  particle loss and gain.  While an algorithmic approach to achieve this balance was previously reported for a single--particle system \cite{Elenewski2015}, numerical experiments demonstrate that this method may be immediately generalized to the multiparticle case without modification.  Interestingly, the contributions from Hartree and exchange correlation effects act to stabilize propagation, expediting the convergence of steady--state currents and enhancing conservation of particle number.  Further improvements are expected when increasing the number of particles and for simulations performed using higher--dimensional systems.

\par There is an important correspondence between this method and other techniques that have been devised to model current carrying devices.  If the density constraint is implemented by copying a mirror image of the wavefunction from the sink region $\mathcal{V}_-$ to the source domain $\mathcal{V}_+$, the  topology of the system  is changed from a closed subinterval $\mathcal{V} \subset \mathbb{R}$ to that of the one--sphere $S^1$.  The external electrical bias is then mapped to an effective external gauge field $A(t)$, similar to previously reported algorithmic constructions \cite{Hubner1996, Gebauer2004}.  In these methods, the electronic wavefunction was periodically multiplied by a phase factor $\exp[-iA(t) x]$ to restore the field configuration to its initial state.  This process is mimicked in our constraint protocol by fixing the phase of the seed density within the generating region to a designated value at each update \cite{Elenewski2015}.

\par A second complication arises from the lack of dissipation due to electron--phonon scattering in our framework.  An electronic excitation will, on average, only propagate as far as a material's mean free path $\ell_p$ before it is scattered.  While this scale is sufficiently long at low temperatures, the associated distance at room temperature is short enough that the scattering time $t_s = \ell_p / v_g$ may be comparable to the simulation timescale, as defined in terms of the mean group velocity $v_g$  for a given state.  If no scattering is present, the simulated currents may exceed those possible in a realistic device. To resolve this, an auxiliary phonon field may be introduced through  the addition of new superoperators to the master equation (Eq. \ref{themastereqn}), thereby introducing the compensatory dissipation needed for a physically accurate simulation.  This approach is well--documented, and has been previously implemented in a manner compatible with our construction \cite{Gebauer2004, Burke2005}.  Interestingly, solutions to a time--dependent KS master equation only  require a nonzero yet arbitrary dissipation parameter to converge a given nonequilibrium steady--state  solution, and hence the precise determination of this coupling strength is not expected to complicate further investigations \cite{Burke2005}.

\par A final consideration must be made when applying Lindblad--like master equations to a Kohn--Sham  mean--field scheme.  While an analogue of the Runge--Gross theorem establishes the uniqueness of the density for a given single--body potential, the rate at which the system approaches a given NESS may not be identical to that of the true physical system \cite{Burke2005}.  Since the mean--field potential assumes the form

\begin{equation}
\mathcal{K}[\rho(x,t)] = \int dx' \frac{\rho(x',t)}{\vert x - x'\vert} + \epsilon_{xc} [\rho(x,t)]
\end{equation} 

\noindent in the continuum limit, any needed corrections may be absorbed into the exchange correlation functional $\epsilon_{xc} [\rho(x,t)]$.  It has been argued that the typical forms of correlation employed in TDDFT, such as the adiabatic local density approximation, are likely to suffice for most practical purposes.  This is particularly true for dissipative effects at weak coupling \cite{Burke2005}.  Whether this consideration applies to genertic particle sources and sinks remains to be seen.

\section{Acknowledgements}
\par This research was partially supported by a start--up grant from The George Washington University (GWU). The authors Y. Zhao and H. Chen also received support from the Dean's Interdisciplinary Collaboration Excellence award at GWU, and Y. Zhao acknowledges additional support from the Simons Foundation through Grant No. 357963.   Computational resources were provided by the Argonne Leadership Computing Facility (ALCF) through the ASCR Leadership Computing Challenge (ALCC) award under Department of Energy Contract DE--AC02--06CH11357 and by the Extreme Science and Engineering Discovery Environment (XSEDE) under National Science Foundation contract TG--CHE130008.

\section{Appendix: RT--TDDFT Algorithm}

During an RT-TDDFT calculation, the time-dependent wavefunction $\psi_{\alpha}$ of the $\alpha$-th
single particle state is obtained by directly propagating solutions to the Schr\"{o}dinger equation (with $\hbar = m_e = e = 1$) forward in time 
\begin{align}
i \dfrac{\partial \psi_{\alpha}(x,t)}{\partial t} = \left( -\dfrac{\nabla^2}{2}  + \hat{V}_{\text{eff}}  \right) \psi_\alpha (x,t)
\end{align}

\noindent where $x\in (x_{\text{L}}, x_{\text{R}})$ is the domain of the simulation and 
$\hat{V}_{\text{eff}} = \hat{V}_{\text{eff}}\big(x,t,\rho(x,t)\big)$ is an effective single-particle, mean-field potential with $\rho(x,t) = \sum_{\alpha = 1}^N \rho_{\alpha}(x,t)$ being the total density. The potential may be succinctly expanded,

\begin{multline}
\hat{V}_{\text{eff}}\Big(x,t,\rho(x,t)\Big) = \int dx' \dfrac{\rho(x',t)}{|x-x'|} + \epsilon_{xc}[\rho(x,t)] \\
 + V_{\text{ext}}(x) + V_{\text{Im}}(x)
\end{multline}

\noindent so that the first term corresponds to the Hartree contribution $V_{\text{H}}(x,t)$, the second term $ \epsilon_{xc}[\rho(x,t)]$ is the exchange correlation, $V_{\text{ext}}(x)$ is the static potential energy landscape, and $V_{\text{Im}}(x)$ contains imaginary source and sink potentials at the simulation boundaries. The Hartree term is evaluated by solving the Poisson equation $\nabla^2 V_{\text{H}}(x,t) = -4\pi \rho(x,t)$ at each time step, and an applied electrostatic bias is introduced by applying inhomogeneous boundary condition to the Poisson solver. In our rudimentary case, the exchange correlation energy may be approximated as $\epsilon_{xc}[\rho(x,t)] = v_x[\rho(x,t)] + v_c[\rho(x,t)]$ where $v_x[\rho(x,t)] = 1(1/\pi)[3\pi^2\rho(x,t)]^{1/3}$ is the Hartree-Fock exchange energy at the Fermi level of a free electron gas. The effects of correlation $v_c[\rho(x,t)] = 0$ are neglected for simplicity \cite{Hohenberg1964, Kohn1965}.

While the wavefunctions at $t$ and $t+\Delta t$ are formally related by the propagator $\psi_{\alpha}(x, t+ \Delta t) = \exp(-i\hat{H}\Delta t)\psi_{\alpha}(x,t)$, a practical implementation requires a numerically efficient scheme to facilitate this evolution. To do so, we define a discrete grid with grid points $(x_j, t_n)$ where
\begin{align}
x_0 = x_{\text{L}}, \quad & x_j = x_{\text{L}} + jh\ (j = 1,2,\cdots, M), \\
\quad x_{M+1} = x_{\text{R}}, & \quad t_n = nk.
\end{align}
\noindent Here $h = \Delta x =  (x_{\text{R}}-x_{\text{L}})/(M+1)$ is the uniform mesh spacing on $x$ and $k = \Delta t$ is the time step. Let $\Psi_{\alpha, j}^n \approx \psi_{\alpha}(x_j,t_n)$ represent the numerical approximation at grid point $(x_j,t_n)$, and introduce $V_j^n = \hat{V}_{\text{eff}}\big(x_j, t_n, \rho(x_j, t_n)\big)$ for notational abbreviation. Furthermore, let $\mathbf{\Psi}_{\alpha}^n, \textbf{V}^n$ be the numerical approximations at $n$-th time step

\begin{align}
\mathbf{\Psi}_{\alpha}^n &= ( \Psi_{\alpha,1}^n,  \Psi_{\alpha,2}^n,\cdots,  \Psi_{\alpha,M}^n )^T, \\
\quad \mathbf{V}^n &= (V_1^n, V_2^n,\cdots, V_M^n)^T.
\end{align}

\noindent Now we first apply the method of line discretization to the  Schr\"{o}dinger equation; namely, we first discretize in space alone, which gives a large system of ODEs with each component of the system corresponding to the solution at some spatial grid point as a function of time. The system of ODEs can then be solved using numerical techniques such as the Runge--Kutta method \cite{LeVeque2007}.

\par For the numerical results herein, we discretize the  Schr\"{o}dinger equation in space at each grid point $x_i$  using a standard centered finite difference scheme:

\begin{multline}
i\Psi_{\alpha,j}'(t) = -s\big( \Psi_{\alpha,j-1}(t) - 2 \Psi_{\alpha,j}(t)  + \Psi_{\alpha,j+1}(t)  \big) \\ 
+ V_j(t) \Psi_{\alpha,j}(t)     \quad \text{for}\ j = 1, 2, \cdots,M,
\end{multline}

\noindent where the prime represents differentiation with respect to time and $s = 1/(2h^2)$. An explicit fourth--order Runge--Kutta method is adopted to integrate the above system of ODEs and ensure numerical stability over a broad range of simulation parameters. The propagation term $\textbf{F}_{\alpha}$ for the Schr\"{o}dinger equation in the presence of an imaginary potential \cite{Elenewski2014}

\begin{align}
\textbf{F}_{\alpha}(\mathbf{\Psi}_{\alpha}^n, \mathbf{V}^n) = [ F_{\alpha,1}, F_{\alpha,2},\cdots F_{\alpha,M} ]^T:  \mathbb{R}^M\rightarrow \mathbb{R}^M
\end{align}
\noindent reads componentwise as

\begin{align}
F_{\alpha,j}(\mathbf{\Psi}_{\alpha}^n, \mathbf{V}^n) = \text{Re}[F_{\alpha,j}] + \text{Im}[F_{\alpha,j}]\quad 
\end{align}
\noindent  for $j = 1, 2, \cdots, M$ with

\begin{equation}
\begin{split}
&\text{Re}[F_{\alpha,j}(\mathbf{\Psi}_{\alpha}^n, \mathbf{V}^n)] =\\  
&-s\Big( \text{Im}(\Psi_{\alpha,j+1}^n) - 2  \text{Im}(\Psi_{\alpha,j}^n) + 
 \text{Im}(\Psi_{\alpha,j-1}^n)\Big) \\  
&+\text{Re}(V_j^n)\text{Im}(\Psi_{\alpha,j}^n) + \text{Im}(V_j^n)\text{Re}(\Psi_{\alpha,j}^n),
\end{split}
\end{equation}
\begin{equation}
\begin{split}
&\text{Im}[F_{\alpha,j}(\mathbf{\Psi}_{\alpha}^n, \mathbf{V}^n)] = \\
&s\Big( \text{Re}(\Psi_{\alpha,j+1}^n) - 2  \text{Re}(\Psi_{\alpha,j}^n) + 
 \text{Re}(\Psi_{\alpha,j-1}^n)\Big) \\  
&+\text{Im}(V_j^n)\text{Im}(\Psi_{\alpha,j}^n) - \text{Re}(V_j^n)\text{Re}(\Psi_{\alpha,j}^n). 
\end{split}
\end{equation}

\noindent A staging procedure is introduced so that the wavefunction increment is evaluated at $t_n, t_{n+1/2}, t_{n+1}$. This is encapsulated through a series of functions
\begin{align}
\mathbf{K}_{\alpha}^1 &= \mathbf{F}_{\alpha}(\mathbf{\Psi}_{\alpha}^n, \mathbf{V}^n),\label{RK01}\\
\mathbf{K}_{\alpha}^2 &= \mathbf{F}_{\alpha}(\mathbf{\Psi}_{\alpha}^n + (k/2)\mathbf{K}_{\alpha}^1, \mathbf{V}^n),\\
\mathbf{K}_{\alpha}^3 &= \mathbf{F}_{\alpha}(\mathbf{\Psi}_{\alpha}^n + (k/2)\mathbf{K}_{\alpha}^2, \mathbf{V}^n),\\
\mathbf{K}_{\alpha}^4 &= \mathbf{F}_{\alpha}(\mathbf{\Psi}_{\alpha}^n + k\mathbf{K}_{\alpha}^3, \mathbf{V}^n),\\
\mathbf{K}_{\alpha} &= \dfrac{1}{6}(\mathbf{K}_{\alpha}^1 + 2\mathbf{K}_{\alpha}^2 + 2\mathbf{K}_{\alpha}^3 + \mathbf{K}_{\alpha}^4),\label{RK05}
\end{align}
\noindent so that the propagated wavefunction at time $t_n$ is given by
\begin{align}
\mathbf{\Psi}_{\alpha}^{n+1} = \mathbf{\Psi}_{\alpha}^n + k \mathbf{K}_{\alpha}.
\end{align}
\noindent Throughout this integration procedure, hard--wall zero Neumann boundary conditions are enforced by setting 

\begin{align}
\Psi_{\alpha,0}^n = \Psi_{\alpha,1}^n, \quad \Psi_{\alpha,M}^n = \Psi_{\alpha,M+1}^n,
\end{align}

\noindent for each wavefunction at each time step $t_n$ of the algorithm.  Note that all quantities comprising the Runge--Kutta integrator (\ref{RK01}-\ref{RK05}) are functions of the same effective potential $V_j^n = \hat{V}_{\text{eff}}\big(x_j, t_n, \rho(x_j, t_n)\big)$, as this mean--field quantity is recalculated only after propagating $\mathbf{\Psi}_{\alpha}^n$ at time step $t_n$ of the simulation. Furthermore, the Hartree potential is determined to high accuracy as a direct Poisson solver is utilized for the simplistic one--dimensional system simulated herein.

The wave propagator can be improved by using spectral differentiation in space, accurate exponential time integrators and  efficient discrete FFT--based algorithms, which will be tested in future work \cite{Ju2015, Ju2015b}.

\bibliography{manuscript}

\end{document}